\documentclass[prl,showpacs,superscriptaddress,twocolumn]{revtex4-1}
\usepackage{graphicx}
\usepackage{epsfig}
\usepackage{amssymb,amsmath}
\usepackage{textcomp}
\usepackage{wasysym}
\usepackage{psfrag}
\usepackage{wrapfig}
\usepackage{color}

\begin{document}

\newcommand{\kv}[0]{\mathbf{k}}
\newcommand{\Rv}[0]{\mathbf{R}}
\newcommand{\rv}[0]{\mathbf{r}}
\newcommand{\al}[0]{\mathbf{a_{1}}}
\newcommand{\as}[0]{\mathbf{a_{2}}}
\newcommand{\K}[0]{\mathbf{K}}
\newcommand{\Kp}[0]{\mathbf{K'}}
\newcommand{\dkv}[0]{\delta\kv}
\newcommand{\dkx}[0]{\delta k_{x}}
\newcommand{\dky}[0]{\delta k_{y}}
\newcommand{\dk}[0]{\delta k}
\newcommand{\cv}[0]{\mathbf{c}}
\newcommand{\qv}[0]{\mathbf{q}}
\newcommand{\Rr}[0]{\Rv_{\rv}}
\newcommand{\Gv}[0]{\mathbf{G}}
\newcommand{\ev}[0]{\mathbf{e}}

\setlength{\jot}{2mm}
\newcommand{\jav}[1]{{\color{red}#1}}

\title{Dissipation induced Luttinger liquid correlations in a one-dimensional Fermi gas}
\author{\'Ad\'am B\'acsi}
\email{bacsi.adam@sze.hu}
\affiliation{MTA-BME Lend\"ulet Topology and Correlation Research Group,
Budapest University of Technology and Economics, 1521 Budapest, Hungary}
\affiliation{Department of Mathematics and Computational Sciences, Sz\'echenyi Istv\'an University, Gy\H or, Hungary}
\author{C\u at\u alin Pa\c scu Moca}
\affiliation{BME-MTA  Exotic  Quantum  Phases  Research Group,   Budapest  University  of  Technology  and  Economics,  1521  Budapest,  Hungary}
\affiliation{Department  of  Physics,  University  of  Oradea,  410087,  Oradea,  Romania}
\affiliation{Department of Theoretical Physics, Budapest University of Technology and Economics, Budapest, Hungary}
\author{Bal\'azs D\'ora}
\affiliation{MTA-BME Lend\"ulet Topology and Correlation Research Group,
Budapest University of Technology and Economics, 1521 Budapest, Hungary}
\affiliation{Department of Theoretical Physics, Budapest University of Technology and Economics, Budapest, Hungary}
\date{\today}

\begin{abstract}
We study a one-dimensional Fermi gas in the presence of dissipative coupling to environment through the Lindblad equation. 
The dissipation involves energy exchange with the environment and favours the relaxation of electrons to excitations.
After switching on the dissipation, the system approaches a steady state, which is described by a generalized Gibbs ensemble.
The fermionic single particle density matrix resembles deceivingly to that in a hermitian interaction quench.
It decays inversely with the distance for short times due to the fermionic 
correlations in the initial state, which changes into a non-integer power law decay for late times, representing dissipation induced Luttinger liquid behaviour.
However, the crossover between the two regions occurs due to dissipation induced damping, and is unrelated to the propagation of excitations.
The velocity of information spreading is set by the dissipative coupling, and differs significantly from the original sound velocity.
The thermodynamic entropy grows as $-t\ln t$ initially, and saturates to an extensive value.
Our results can be tested experimentally in one-dimensional Dirac systems.

\end{abstract}

\maketitle
%\tableofcontents

%%%%%%%%%%%%%%%%%%%%%%%%%%%%%%%%%%%%%%%%
\paragraph{Introduction.} 
Thanks to the advent of sophisticated experimental technologies in cold atomic settings and in condensed matter, 
the creation and controlled manipulation of isolated quantum systems became possible\cite{polkovnikovrmp,dziarmagareview,BlochDalibardZwerger_RMP08}.
In particular, one can follow the spatio-temporal dynamics of strongly interacting quantum gases\cite{gring,erne} after some arbitrary time evolution protocol.
The emerging universal behaviour and scaling provides not only essential information on the (pre-)thermalization and relaxation, 
but is also relevant to simulate the  early Universe after inflation,
for which the experimental knobs are obviously limited.
All this information becomes relevant for quantum computation and information processing\cite{nielsen}.

However, no system is perfectly isolated from the environment, therefore considering open quantum systems, coupled to some external bath is necessary
to understand realistic systems.
In its simplest form, this is taken into account by the Lindblad equation\cite{daley,carmichael,fujii}. This enterprise already 
gives way to engineer peculiar, dissipation induced states of matter
with no obvious analogues in closed quantum systems\cite{diehl,pichler,barreiro,buca,naghiloo,nhkitaev2018,ashidaprl2}.

For closed quantum systems, Landau's Fermi liquid picture provides a good description of the normal state of many interacting systems in dimensions higher than one\cite{abrikosov}. Therein, many properties of the original Fermi gas are inherited,
though certain properties are renormalized. This picture breaks down in one dimension, and the ensuing interacting ground state differs markedly from that of the initial Fermi gas\cite{nersesyan,giamarchi}.
The original fermionic excitations are replaced by bosonic collective modes, consisting of many electron-hole pairs.
Given the apparent vulnerability of a one-dimensional Fermi gas in closed quantum systems to Luttinger liquid or gap opening instabilities\cite{nersesyan,giamarchi}, 
their fate in an open quantum system is still an open question, i.e. when the fermionic degrees of freedom living in one dimension are coupled dissipatively to some environment.
In particular, what is the structure of the ensuing steady state and what characterizes the non-unitary dynamical evolution towards the steady state?

This motivated us to investigate a one-dimensional Fermi gas in the presence of dissipative coupling to environment through the Lindblad equation.
The dissipation involves energy exchange with the environment and favours the relaxation of electrons to excitations.
We follow the full non-unitary dynamics of the system after switching on the dissipation at $t=0$. Other systems were also investigated in similar context\cite{Keck,ashida18}.
We find that the steady state is described \emph{exactly} by a generalized Gibbs ensemble.
The dissipation induces Luttinger liquid like correlation during the time evolution, but the velocity of information spreading is set by the dissipative coupling, and is unrelated to the sound velocity.
Our findings can be tested with current experimental technologies.

\paragraph{Dissipation in a one-dimensional Fermi gas.} 
We consider non-interacting one-dimensional spinless electrons,
 which, within the realm of a low energy theory, can propagate to left or right\cite{giamarchi,nersesyan}.
The low energy effective theory of the electrons in bosonized form gives rise to the Luttinger model with the Hamiltonian
\begin{gather}
H=\sum_{q>0}\omega_0(q)\left(b^{+}_qb_q + b^{+}_{-q}b_{-q}\right)
\label{eq:ham}
\end{gather}
where $b_q$ is the annihilation operator of the bosonic excitations which is bilinear of the original 
fermionic operators\cite{giamarchi,delft,nersesyan}. In Eq. \eqref{eq:ham},
$\omega_0=v|q|$ is the non-interacting spectrum with $v$ the sound (or Fermi) velocity, describing low energy excitations around $\pm k_F$ with $k_F$ the Fermi momentum.

\begin{figure}[h!]
\centering
\includegraphics[width=8.5cm]{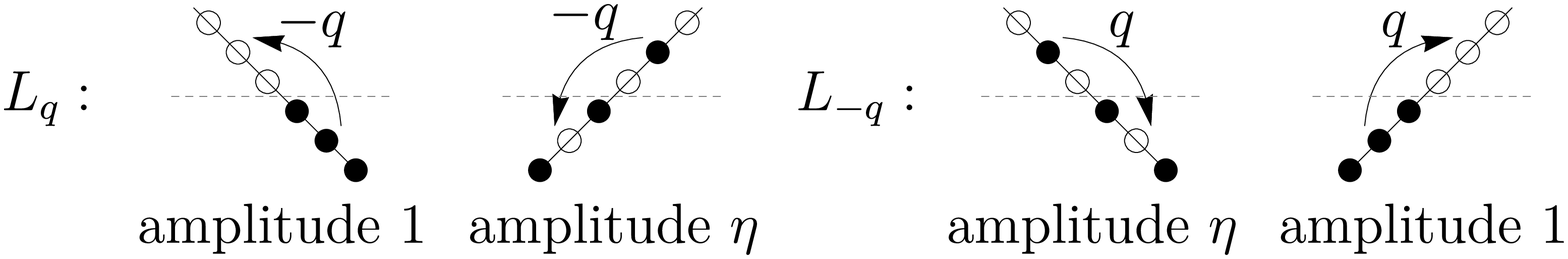}
\caption{Illustration of jump operators in the Lindblad equation, $L_q$ and $L_{-q}$ create electron-hole pairs on the right (around $k_F$) 
and left (around -$k_F$) moving branches with momentum $-q$ and $q$, respectively.  The dashed line denotes the Fermi energy,
filled/empty circles stand for electron/holes, respectively.
The jump operators are mixtures of excitation and relaxation of an electron with amplitude $1$ and $\eta$, respectively.}
\label{fig:jumpop}
\end{figure}

In an open quantum system, coupling to  environment induces non-unitary time evolution which is described by the Lindblad equation as
\begin{gather}
\partial_t \rho = -i\left[H,\rho\right]+\sum_{q\neq 0}\left(\left[L_q,\rho L_q^+\right]  + h.c.\right),
%\partial_t \rho=-i\left({H}_{\rm eff}{\rho}-{\rho}{H}_{\rm eff}^{\dagger}\right)+2\gamma\sum_{q\neq 0}{L}_{q}{\rho}{L}^{\dagger}_{q},
\label{eq:lindblad}
\end{gather}
which determines the dynamics of the density matrix $\rho(t)$.
This dissipative coupling to environment  is taken into account by the jump operators $L_q$, which are best introduced visually in Fig. \ref{fig:jumpop} for our current system.
They  only involve energy exchange with the environment and favour the relaxation of electrons to excitations. Similar jump operators were considered in Ref. \onlinecite{diehl}.
The bosonized jump operators, visualized in Fig. \ref{fig:jumpop} are
\begin{gather}
L_{q}=\sqrt{\gamma |q|}\left(\eta b_q + b_{-q}^+\right)
\label{eq:jumpop}
\end{gather}
with $\eta>0$ \footnote{We note that in the case of a negative or complex $\eta$, the complex phase could be scaled out by redefining the $b_q$ operators.}, and
 $\gamma$  measures the strength of the coupling between the system and the environment and has velocity dimension.
With this choice of jump operators, the ensuing problem becomes genuinely many-body.

In general, the jump operators of the Lindblad equation describe the elementary processes occurring while the system interacts with its environment. The operators in
Eq. \eqref{eq:jumpop} are chosen in such a way that they describe electron-hole excitation and relaxation while the total momentum of the system is shifted by 
momentum $-q$. Electron excitations increase the system energy with $\omega_0(q)$ while electron relaxation decreases it with the same amount. 
These processes are taken into account with different amplitudes and the parameter $\eta$ enables us to describe either 
dissipation of energy to the environment or energy pumped into the system. 
Our choice of jump operators is also motivated by the possibility
of studying dissipative effects analytically, focusing on features that do not depend qualitatively on the form of the coupling to the environment.
Furthermore, the operator in Eq. \eqref{eq:jumpop} can be regarded a generalization of the electron density since 
 $L_q$ is proportional to the Fourier transformed electron density for $\eta=1$.
This limit was considered in Ref. \cite{medvedyeva16,buchhold2014}.

To set the stage and to appreciate the role of $\eta$, we first calculate the time evolution of the average number of excitations, $\hat{n}_q=b_q^+b_q$, as
\begin{gather}
n_q(t)=\mathrm{Tr}\left[\rho(t)\hat{n}_q\right]=\nonumber \\
=\frac{1}{\eta^2-1} + \left(n_q(0) - \frac{1}{\eta^2-1}\right) e^{-2\gamma |q| t (\eta^2-1)}
\label{eq:occ}
\end{gather}
where $n_q(0)$ is the occupation number in the initial state.
For $\eta>1$, i.e. when the boson annihilation has a larger amplitude compared to the boson creation, 
the boson number relaxes to $1/(\eta^2-1)$. This indicates that the system has a stable steady state. 
For $\eta\le 1$, however, the occupation number explodes and the system is essentially boiled up to infinite temperatures. 

%For the special case of $\eta=1$, the occupation number increases linearly as $n_q(t) = n_q(0) + 2\gamma |q| t$ 
%in accordance with \cite{buchhold2014}. This also corresponds to a boiled up system.

By studying Eq. \eqref{eq:lindblad}, it is remarkable that the Lindblad equation only couples $q$ and $-q$ modes 
as long as the initial state does not couple additional modes.
This allows us to write $\rho(t)=\prod_{q>0}\rho_q(t)$.

\paragraph{Time evolution and steady state of the Lindblad equation.} 
For one specific $q>0$ mode, the solution of Eq. \eqref{eq:lindblad} is assumed in the form of
\begin{gather}
\rho_q(t)=r_q(t)e^{c_q(t)b_{q}b_{-q}}e^{-\ln(\nu_q(t)+1)\left(b_q^+b_q + b_{-q}b_{-q}^+\right)}\times\nonumber\\
\times e^{c_q(t)^*b_q^+b_{-q}^+}
\label{eq:rhodef}
\end{gather}
where $\nu_q(t)$ and $r_q(t)$ are real functions of time and $c_q(t)$ is a complex-valued function.
The trace of the density matrix is preserved when $\nu_q(t)>0$ and 
\begin{gather}
r_q(t)=\frac{\nu_q(t)^2-|c_q(t)|^2}{\nu_q(t)+1}>0
\label{eq:rdef}
\end{gather}
at any time instant. The latter equality shows that $r_q(t)$ is expressed with $\nu_q(t)$ and $c_q(t)$, therefore, the density matrix is completely characterized by these two functions.
The average number of excitations is written as $n_q(t) = \nu_q(t)/(\nu_q(t)^2-|c_q(t)|^2)$
which has already been evaluated in Eq. \eqref{eq:occ}.

By substituting Eq. \eqref{eq:rhodef} into the Lindblad equation Eq. \eqref{eq:lindblad}, we obtain after some lengthy algebra\cite{EPAPS}
\begin{subequations}
\begin{gather}
\dot{\nu}_q=-2\gamma |q| \Big(|c_q|^2 + \nu_q^2 + \nu_q\left(1-\eta^2\right) + \nu_q \eta \left(c_q+c_q^*\right)\Big)  \label{diff1}\\
\dot{c}_q = 2iv|q|c_q+2\gamma |q| \Big(c_q\left(\eta^2-1\right)-\eta\left(\nu_q^2+c_q^2\right)-2\nu_q c_q\Big).
\end{gather}
\label{eq:diffeq}
\end{subequations}
%and the initial condition, corresponding to the ground state at $T=0$, is $\nu_q(0)=\infty$, $c_q(0)=0$.
Despite of their non-linear nature, the differential equations can be solved and the stable steady states can be determined analytically \cite{EPAPS}.
For $\eta\leq 1$, the stable steady state is 
$\nu_{\rm ex} = 0$ and $c_{\rm ex}=0$. For these values, however,
no density matrix can be assigned since $\nu_{\rm ex}$ is out of the domain $\nu>0$. Nevertheless, the steady state can be interpreted physically as the boiled up system which is characterized by an infinite temperature.
This is in accordance with the preliminary calculations of the occupation number. Namely, for $\eta\leq 1$, boson annihilation (electron relaxation) is not strong enough to damp the system, and the environment induces energy explosion. 

For $\eta>1$, when boson annihilation is expected to be strong enough to realize energy dissipation in the system, the stable fix point of the differential equations is 
\begin{gather}
\nu_{\infty} =|A|^2 \frac{\eta^2-1}{|A|^2-\eta^2}\qquad\qquad c_{\infty}=-\frac{\nu_\infty\eta}{|A|^2}A
\label{eq:sol2}
\end{gather}
with
$A=1 + \frac{iv}{\gamma(\eta^2-1)}$.
Neither $A$ nor the steady parameters $\nu_\infty$ and $c_\infty$ depend on the wavenumber, therefore the stationary density matrix is the same in all $q>0$ channels.
This stationary density matrix is rewritten as
\begin{gather}
\rho_q(t\rightarrow \infty; \eta>1) = (1-e^{-\Omega_\infty})^2e^{-\Omega_\infty\left( d^+_{q} d_{q} + d_{-q}^{+}d_{-q}\right)}
\label{eq:rhodiag}
\end{gather}
where
%\begin{gather}
%\Omega_\infty = \left|\mathrm{arcosh}\left(\frac{\nu_\infty^2-|c_\infty|^2}{2(\nu_\infty+1)}+1\right)\right|
%\end{gather} 
$\Omega_\infty = |\textmd{acosh}((\nu_\infty^2-|c_\infty|^2)/(2(\nu_\infty+1))+1)|$
is independent from the wavenumber $q>0$. The operators $d_q$ describe the eigenstates 
of the steady state and can be calculated via Bogoliubov transformation\cite{EPAPS}. 
The steady density matrix resembles a thermal state with 
$\omega_0(q)/T=\Omega_\infty$. The wavenumber independence of $\Omega_\infty$ implies 
that the temperature must depend on the momentum as $T(q)\sim |q|$. Therefore, the overall 
steady state density matrix describes exactly a generalized Gibbs ensemble (GGE)\cite{rigol}
~\footnote{The Lindblad equation only yields a conventional thermal steady state density matrix when the couplings are fine tuned to satisfy detailed balance of a 
canonical ensemble\cite{breuer}.}.
Note that a similar density matrix describes only approximately the steady state of a 
Luttinger liquid after a hermitian interaction quench\cite{cazalillaprl,dorapdf}.

\paragraph{Single particle density matrix.}
The density matrix enables us to calculate various physical quantities. We start with its single particle version, which is related to the Green's function in equilibrium.
Since the fermion field decomposes to right and left moving parts as $\Psi(x)=e^{ik_F x}\Psi_R(x) +e^{-ik_F x}\Psi_L(x)$, it is enough\cite{giamarchi,cazalillaprl} to concentrate
on
\begin{gather}
G(x;t)=-i\mathrm{Tr}\left[\rho(t)\Psi_R^+(x)\Psi_R(0) \right]
\label{eq:greendef}
\end{gather}
where $\Psi_R(x)$ is the fermionic field operator of right-moving electrons with $\Psi_R(x)=\frac{1}{\sqrt{2\pi\alpha}}
\exp\left[i\sum_{q>0}\sqrt{\frac{2\pi}{qL}}\left(e^{iqx}b_q+e^{-iqx}b^+_q \right)\right]$, describing excitations  around the right Fermi momentum $k_F$.
By following standard steps\cite{giamarchi,delft}, we obtain
\begin{gather}
\frac{G(x;t)}{G_0(x)}=
%-\sum_{q>0}\frac{4\pi}{Lq}\frac{\nu_q(t)}{\nu_q^2(t)-|c_q^2(t)|}\left(1-\cos(qx)\right)=\nonumber \\
\exp\left(-\sum_{q>0}\frac{4\pi}{Lq}n_q(t)\left(1-\cos(qx)\right)\right),
\label{eq:green3}
\end{gather}
where $L$ is the system size~\footnote{Similar expression stands for the correlator of the left movers as well. 
Altogether, the Green's function of the physical $\Psi$ fermion field is $2i$Im$[e^{ik_F x}G(x;t)]$.}. It is remarkable that all the time-dependence of the single particle density matrix occurs only through the average number of excitations.
The function $G_0(x)=1/(x+i \alpha)2\pi$ is the correlation function of non-interacting fermions at zero temperature. 
The length scale $\alpha$ is in the range of the lattice constant and is introduced as an exponential cut-off in momentum space, $\exp(-\alpha q)$.

The time-dependence of $n_q(t)$ is already obtained in Eq. \eqref{eq:occ}. By starting initially from the ground state of the non-interacting Fermi gas, no excitations are present and $n_q(0)=0$. 
In the thermodynamic limit, we obtain from Eq. \eqref{eq:green3}
\begin{gather}
\ln \frac{G(x;t)}{G_0(x)}=\frac{1}{1-\eta^2}\ln\left(\frac{1+\left(\frac{x}{\alpha}\right)^2}{1+\left(\frac{x}{\alpha+2\gamma t(\eta^2-1)}\right)^2}\right).
\label{eq:grf}
\end{gather}
The most notable feature in Eq. \eqref{eq:grf} is that the time evolution is governed by the speed $\gamma(\eta^2-1)$, which can differ significantly from the
original sound velocity $v$. For unitary time evolution (i.e. in the absence of dissipative coupling), any time dependence would be
dictated by (a renormalized) $v$. For the Lindblad equation, however, the eigenvalues of the r.h.s of Eq. \eqref{eq:lindblad} have negative real part\cite{daley} (except for the steady state),
whose magnitude is controlled by $\gamma$. After switching on the dissipation, any transient component of the density matrix dies out during the time evolution
exactly due to the presence of $\gamma$. Therefore, the
velocity of information spreading is set by the dissipative coupling, and differs from the original sound velocity.

\begin{figure}
\centering
\includegraphics[width=8cm]{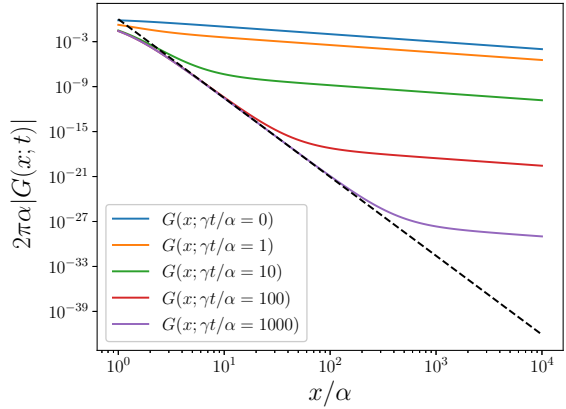}
\caption{The spatial dependence of the single particle density matrix,  Eq. \eqref{eq:grf} for several times and $\eta=1.3$. 
Initially at $\gamma t=0$, it decays as $x^{-1}$ (blue line). During the time evolution, the 
correlation function turns to the power-law decay as $x^{ - \frac{\eta^2+1}{\eta^2-1}}$ for $x\ll \gamma t(\eta^2-1)$, and retains the
$x^{-1}$ decay for $x\gg\gamma t(\eta^2-1)$. The black dashed line denotes the steady state behaviour.}
\label{fig:grf}
\end{figure}

At $t=0$, the right-hand side of Eq. \eqref{eq:grf} vanishes and the correlation function is just equal to $G_0(x)$.
After switching on the dissipative coupling at $t=0$, the initial $G(x;t=0)\sim x^{-1}$ correlations are still retained for short times.
Indeed, for $x\gg\gamma t(\eta^2-1)$, the Green's function gets dressed with a time dependent quasiparticle residue,
$Z(t)=\left(1+2\gamma t(\eta^2-1)/\alpha\right)^{-2/(\eta^2-1)}$, which decays as a power law of time.
This indicates that due to dissipation, the initial non-interacting state gets renormalized and heavy fermionic.
On the other hand, for $x\ll\gamma t(\eta^2-1)$, the fermionic nature of quasiparticles is lost and gives way to a non-integer Luttinger liquid like exponent, summarized as
\begin{gather}
G(x;t)\sim\left\{\begin{array}{lc}
\dfrac{Z(t)}{x} & \textmd{ for }x\gg\gamma t(\eta^2-1)\\
\left({x}/{\alpha}\right)^{ - \frac{\eta^2+1}{\eta^2-1}} & \textmd{ for } x\ll\gamma t(\eta^2-1)
\end{array}\right..
\label{greenlimits}
\end{gather}
These features are highlighted in Fig. \ref{fig:grf}, indicating a smooth transition between the short and long distance decay.
%Our calculation also allows to address the $\eta=1$ case\cite{buchhold2014}, when the steady state heats up to infinite temperatures. 
%The Green's function decays exponentially as $G(x\gg\alpha;t)\sim \exp(-4\gamma t/\alpha)/x$.

The late time power-law  
exponent is $-(\eta^2+1)/(\eta^2-1)$. Similar non-integer exponents are familiar for Luttinger liquids,
where the equilibrium exponent is well known\cite{giamarchi} to be $-(K+K^{-1})/2$ at $T=0$ with $K$ being the Luttinger parameter. In our setting, however,
no electron interaction is present and the non-trivial exponent occurs solely due to the dissipation.
It is important to note that the coupling to the environment does not lead to an effectively interacting \emph{ground state} in the long time limit.
On the contrary, we found that the steady density matrix rather represents a GGE which is a thermal state in general sense.

To elaborate on this a bit more, we consider the $2k_F$ oscillating part of the fermionic density correlation function\cite{giamarchi,EPAPS}.
For short times cf. Eq. \eqref{greenlimits}, it retains its original fermionic spatial decay as $\sim \cos(2k_F x)/x^2 t^{\delta-2}$, while
in the long time limit, it exhibits LL behaviour in terms of non-integer spatial power law decay as $\sim \cos(2k_F x)/x^\delta$ with 
$\delta=2\frac{\eta^2+1}{\eta^2-1} - 4 \frac{\eta(\eta^2-1)}{\left(\frac{v}{\gamma}\right)^2+\left(\eta^2-1\right)^2}>2$ \cite{EPAPS}. 
Here, both characteristics of coupling to the environment, $\gamma$ and $\eta$ appear, while the single particle density matrix 
features only $\eta$. This indicates that the conventional LL relations
between various exponents\cite{giamarchi,iucci} do \emph{not} hold for the dissipation induced non-trivial steady state.
 
%The signature of this thermal behavior can be found in the density-density correlation function whose short-range oscillating part \cite{giamarchi} is evaluated as
%\begin{gather}
%C^\mathrm{osc}(x;t)=\frac{2\cos(2k_F x)}{(2\pi\alpha)^2}\left\{\begin{array}{lc}
%\left(\dfrac{\alpha}{x}\right)^2 Z_C(t) & \textmd{ for }x\gg\gamma t(\eta^2-1)\\
%\left(\dfrac{\alpha}{x}\right)^{h} & \textmd{ for } x\ll\gamma t(\eta^2-1)
%\end{array}\right..
%\end{gather}
%where $Z_C(t)=\left(\alpha/(2\gamma t(\eta^2-1))\right)^{h-2}$ and the non-trivial exponent of
%\begin{gather}
%h=2\frac{\eta^2+1}{\eta^2-1} - 4 \frac{\eta(\eta^2-1)}{\left(\frac{v}{\gamma}\right)^2+\left(\eta^2-1\right)^2}
%\end{gather}

\paragraph{Entropy.}
The relaxation toward the steady state is manifested also in the time evolution of the thermodynamic entropy, i.e. $S(t)=-\textmd{Tr}[\rho(t)\ln\rho(t)]$. 
At $t=0$, the system is in a pure state with zero entropy. After switching on the coupling to the environment, the entropy varies with time as
\begin{gather}
S(t)= \sum_{q>0} \frac{2\Omega_q(t)}{e^{\Omega_q(t)}-1}-2 \ln\left(1-e^{-\Omega_q(t)}\right) 
\label{eq:entropy}
\end{gather}
due to the non-unitary nature of the Lindblad equation. In Eq. \eqref{eq:entropy}, $\Omega_q(t)$ is the instantaneous eigenvalue of the exponent in the density matrix\cite{EPAPS}.
For weak dissipation, i.e. when $\gamma\ll v$ and $c_q(t)\approx 0$ is assumed in each momentum channel, the short time growth  ($\gamma t\ll \alpha$) of the entropy is  $S(t)\sim -L \gamma t\alpha^{-2}\ln(\gamma t/\alpha)$.
The entropy satisfies a volume law and keeps growing for $\gamma t \sim \alpha$ and saturates to its steady value afterwards, which follows from substituting $\Omega_\infty$ into Eq. \eqref{eq:entropy}.

The single particle density matrix of our model exhibits similar time-dependence to the evolution
found after a quantum quench in the Luttinger model where the interaction was switched 
on suddenly \cite{cazalillaprl,iucci}, follows by unitary time evolution.
%After the quench, the unitary time-evolution resulted in a similar steady state and single particle density matrix.
%However, 
Despite the similarities in the correlations, there are three essential differences: first of all, 
the velocity of information spreading in our case stems from the decay rate from Lindblad dynamics, while it originates from the propagation of quasiparticle
excitations for the hermitian case and equals to the effective speed of light\cite{liebrobinson}.
Second, the GGE is exact for the present dissipative system and is only approximate for the 
hermitian quantum quench\cite{iucci,dorapdf}.
Finally, dissipation leads to entropy production in our model while unitary time evolution does not change the entropy of the system.

\paragraph{Relation to experiments.}
Experimentally, our setup can be realized by two coupled Luttinger liquids\cite{perfetto2006}, 
interacting through chiral density-density interaction and without electron tunneling.
One Luttinger liquid would represent the bath, which would directly induce the jump operators in Eq. \eqref{eq:jumpop} in the liquid, with $\gamma\eta^2$ and $\gamma$ determined
by the interaction between electron densities of the same and opposite chirality.
In another setting, the jump operators can be implemented in a
 controlled fashion using a lattice realization of the Creutz ladder\cite{viyuela,creutz}, which can also be realized
experimentally\cite{kang19}. When tuned to its critical point, it realizes one-dimensional Dirac fermions\cite{sticlet}, and
 two legs of the ladder\cite{sticlet} host the right and left moving excitations, which are then also spatially separated. This allows for
coupling the right and left moving densities to the environment independently to realize the dissipators depicted in Fig. \ref{fig:jumpop}.
The unequal weights of the $\pm q$ processes in a given branch in Fig. \ref{fig:jumpop} follow naturally, e.g.  from the detailed balance\cite{rajagopal}.

\paragraph{Conclusion.}
We have investigated the fate of a one-dimensional Fermi gas of electrons coupled to a dissipative environment via the Lindblad equation. 
Using Abelian bosonization, the ensuing Lindblad dynamics is solved within the realm of the low energy effective theory.
The steady state density matrix coincides with that of a generalized Gibbs ensemble.
After switching on dissipation suddenly at $t=0$, the single particle density matrix or Green's function exhibits similar spatio-temporal pattern than
 after a hermitian interaction quench\cite{cazalillaprl}. 
This resemblance is, however, deceiving. Due to dissipation, correlations do not propagate
with the effective sound velocity, but are damped by the dissipation, resulting in a significantly different velocity of information spreading.
In addition, the characteristic features of Luttinger liquid correlation in terms of non-integer power law exponents
for the spatial and temporal decay are revealed, but in this case these are induced by dissipation and not by electron-electron interaction.
The thermodynamic entropy grows as $-t\ln t$ initially before saturating to its steady state value, and satisfies a volume law.
These features can be observed in coupled Luttinger liquids or in one-dimensional Dirac systems.

\begin{acknowledgments}
This research is supported by the National Research, Development and Innovation Office - NKFIH   within the Quantum Technology National Excellence Program (Project No.
      2017-1.2.1-NKP-2017-00001), K119442 and by
 a grant from the Simons Foundation.
This work was performed in part at Aspen Center for Physics, which is supported by National Science Foundation grant PHY-1607611.
\end{acknowledgments}

\bibliographystyle{apsrev}
\bibliography{lindblad,wboson1}

\section{Supplementary Material}
\subsection{Time evolution of the density matrix}
In this section, the index $q$ is dropped and all quantities refer to one specific wavenumber $q>0$.
The annihilation operators are henceforth denoted by $b_+$ for positive $+q$ and $b_-$ for the negative momentum $-q$.
The density matrix of Eq. (5) in the main text may be rewritten as
\begin{gather}
\rho(t)=R^+(t)R(t) 
\label{eq:rho2}
\end{gather}
with
\begin{gather}
R(t)=\sqrt{r(t)}e^{-\ln(\nu(t)+1)K_0}e^{c(t)^*K_+}
\end{gather}
where
$K_0=\left(b_{+}^{+}b_{+} + b_-b_-^+\right)/2$ and $K_+=b^+_+b_-^+$
obey the relations of an $su(1,1)$ algebra together with $K_-= K_+^+$. We substitute
 Eq. \eqref{eq:rho2} into the Lindblad equation as given in Eq. (2) in the main text. We use the commutation rules of the $K$ operators and the commutations of $K$ with $b_\pm$ in such a way that all terms are written in the form of $R(t)^+\left(\dots\right)R(t)$. For example, the time derivative is calculated as
\begin{gather}
\partial_t\rho=R^{+}\left(\frac{\dot{r}}{r}+2\frac{\dot{\nu}K_0}{\nu+1}+ \frac{\dot{c}^*K_{+}+\dot{c}K_{-}}{\nu+1}\right)R
\label{eq:drhodt}
\end{gather}
After rearranging all terms to this form, the expressions between $R(t)^+$ and $R(t)$ are all linear combinations of the identity operator and the three operators $K_0$, $K_+$ and $K_-$. For each four operators, the coefficients must equal on both sides of the equation which provides us with four differential equations. The four differential equations mean only two independent equations, since those of $K_+$ and $K_-$ are complex conjugates and $r(t)$ is related to $\nu(t)$ and $c(t)$ by Eq. (6) in the main text. The equations from the coefficients of $K_0$ and $K_-$ are then given by
\begin{subequations}
\begin{gather}
\dot{\nu}=-2\gamma |q| \Big(|c|^2 + \nu^2 + \nu\left(1-\eta^2\right) + \nu \eta \left(c+c^*\right)\Big)  \label{diff2}\\
\dot{c} = 2iv|q|c+2\gamma |q| \Big(c\left(\eta^2-1\right)-\eta\left(\nu^2+c^2\right)-2\nu c\Big)
\end{gather}
\label{eq:diffeq2}
\end{subequations}

The differential equations are analytically solvable by defining the expectation values $n(t)=\mathrm{Tr}[\rho(t)b_\pm^{+}b_\pm]$ and $m(t)= \mathrm{Tr}[\rho(t)b_+^{+}b_-^{+}]$ which are related to the functions $\nu(t)$ and $c(t)$ by
\begin{gather}
n(t) = \frac{\nu(t)}{\nu(t)^2-|c(t)|^2}\qquad\quad m(t)=\frac{c(t)}{\nu(t)^{2}-|c(t)|^2}\,.
\end{gather}
Based on Eqs. \eqref{eq:diffeq2}, the time evolution of these quantities is obtained as
\begin{subequations}
\begin{gather}
\dot{n}=2\gamma|q|\left(1-(\eta^2-1)n\right) \label{diff3} \\
\dot{m}=2\gamma|q|\left(-\eta+\left(i\frac{v}{\gamma} - (\eta^2-1)\right)m\right)
\end{gather}
\label{eq:diffeq3}
\end{subequations}
which are solved by
\begin{subequations}
\begin{gather}
n(t)=\frac{1}{\eta^2-1}+\left(n_0-\frac{1}{\eta^2-1}\right)e^{-2\gamma|q|(\eta^2-1)t}\\
m(t)=\frac{\eta}{i\frac{v}{\gamma} - (\eta^2-1)} + \nonumber \\
+ \left(m_0-\frac{\eta}{i\frac{v}{\gamma} - (\eta^2-1)}\right)e^{2iv|q|t}e^{- 2\gamma|q|(\eta^2-1)t}
\end{gather}
\label{eq:nm}
\end{subequations}
where $n_0$ and $m_0$ are the initial conditions for $n(t)$ and $m(t)$. For a zero temperature initial condition, $n_0=0$ and $m_0=0$. The functions $\nu(t)$ and $c(t)$ can be computed based on Eqs. \eqref{eq:nm} by
\begin{gather}
\nu(t) = \frac{n(t)}{n(t)^2-|m(t)|^2}\qquad\quad c(t)=\frac{m(t)}{n(t)^{2}-|m(t)|^2}
\end{gather}
which describe the time evolution of the density matrix.

For $\eta\leq 1$, the stable steady state of the density matrix is determined by
$\nu_{\rm ex} = 0$ and $c_{\rm ex}=0$.

For $\eta>1$, the stable steady state is characterized by
\begin{gather}
\nu_{\infty} = \frac{\eta^2-1}{1-\frac{\eta^2}{|A|^2}}\qquad\qquad c_{\infty}=-\frac{\nu_\infty\eta}{|A|^2}A
\label{eq:sol3}
\end{gather}
where $A=1 + \frac{iv}{\gamma(\eta^2-1)}$.
It can be observed that 
$\nu_{\infty}$ diverges as $\eta$ reaches $\eta_{c}=\sqrt{1+(v/\gamma)^{2/3}}$. Diverging $\nu_q\rightarrow \infty$ means that the system is perfectly damped and relaxes to the ground state. Beyond the limit, i.e., for $\eta>\eta_c$, 
we obtain $\nu_\infty<0$ and the system is most probably in an impossible squeezed state \cite{fisher1984}.

\subsection{Diagonalization of the density matrix}

The density matrix as given in Eq. (5) of the main text can be diagonalized in two steps. First, one has to rewrite the product of three exponentials in a single exponential by using the commutation rules of the $su(1,1)$ algebra \cite{solomon,gilmore}.
\begin{gather}
\rho(t)=r(t)\,\exp\left(\frac{\left(s(t) K_{-}+2K_0 + s(t)^*K_{+}\right)\Omega(t)}{\sqrt{1-|s(t)|^2}}\right)
\end{gather}
 where
\begin{gather}
\Omega(t) = \left|\textmd{acosh}\left(\frac{\nu(t)^2 - |c(t)|^2}{2(\nu(t)+1)}+1\right)\right|
\label{eq:omegatdef}
\end{gather}
and 
\begin{gather}
s(t) = \frac{2c(t)}{|c(t)|^2-\nu(t)^2-2\nu(t)}\,.
\label{eq:sdef}
\end{gather}
Note that the exponent is quadratic in the bosonic annihilation and creation operators. The second step toward diagonalization is the Bogoliubov transformation
\begin{gather}
\left[\begin{array}{c} d_+(t) \\ d_-(t)^{+}\end{array}\right] = \left[\begin{array}{cc} u(t) & v(t) \\ v(t)^{*} & u(t) \end{array}\right]
\left[\begin{array}{c} b_+ \\ b_-^{+}\end{array}\right]
\label{eq:bogoliubov}
\end{gather}
where
\begin{gather}
u(t)=\frac{1}{\sqrt{2}}\sqrt{\frac{1}{\sqrt{1-|s(t)|^2}}+1} \\
v(t)=\frac{s(t)^{*}}{\sqrt{2}|s(t)|}\sqrt{\frac{1}{\sqrt{1-|s(t)|^2}}-1}\,.
\end{gather}
The transformation leads to
\begin{gather}
\rho(t) = (1-e^{-\Omega(t)})^2e^{-\Omega(t)\left( d_{+}(t)^+ d_{+}(t) + d_{-}(t)^{+}d_{-}(t)\right)}
\end{gather}
which converges to the thermal state of Eq. (9) of the main text in the long time limit.

\subsection{Oscillating part of the fermionic density correlation function}

The short wavelength, $2k_F$ oscillating part of the fermionic density is written in term of bosons
as\cite{giamarchi,iucci}
\begin{gather}
n_{\pm 2k_F}(x)=\frac{1}{2\pi\alpha}\exp\left[\pm\left( 2ik_F-2i\Phi(x)\right)\right],
\end{gather}
where $k_F$ is the Fermi momentum  and
\begin{gather}
\Phi(x)=-i\sum_{q\neq 0}\sqrt{\frac{\pi|q|}{2 L}}\frac{1}{q}e^{-iqx}\left(b_q^{+}+b_{-q}\right)\,.
\end{gather}
We consider the corresponding density-density correlation function\cite{giamarchi} defined as
\begin{gather}
C^{\rm osc}(x;t)=\mathrm{Tr}\left[\rho(t)n_{2k_F}(x)n_{-2k_F}(0)\right] + c.c.
\end{gather}

%We consider the oscillating part of the fermionic density correlation function \cite{giamarchi} defined as
%\begin{gather}
%C^{\rm osc}(x;t)=\frac{e^{-2ik_F x}}{(2\pi\alpha)^2}\mathrm{Tr}\left[\rho(t)e^{2i(\Phi(x)-\Phi(0))}  \right] + c.c.
%\end{gather}
%where $k_F$ is the Fermi momentum  and
%\begin{gather}
%\Phi(x)=-i\sum_{q\neq 0}\sqrt{\frac{\pi|q|}{2 L}}\frac{1}{q}e^{-iqx}\left(b_q^{+}+b_{-q}\right)\,.
%\end{gather}
By evaluating the trace, we obtain
\begin{gather}
\ln\frac{C^{\rm osc}(x;t)}{C^{\rm osc}_0(x)}=-\sum_{q>0}\frac{8\pi}{L|q|}\left(n_q(t) + \mathrm{Re}\,m_q(t)\right)\times \nonumber \\ \times\left(1-\cos(qx)\right)
\label{eq:corr}
\end{gather}
where
\begin{gather}
C^{\rm osc}_0(x)=\frac{2\cos(2k_F x)}{(2\pi)^2}\frac{1}{\alpha^2+x^2}
\end{gather}
is the correlation function in the initial $T=0$ state, exhibiting the characteristic $x^{-2}$ decay at large distances. 
The functions $n_q(t)$ and $m_q(t)$ are given in Eqs. \eqref{eq:nm} with $n_0=0$ and $m_0=0$. The integral in Eq. \eqref{eq:corr} can be calculated analytically. 
In the short time limit, $\alpha\ll\gamma t(\eta^2-1)\ll x$, the correlation function is evaluated as
\begin{gather}
C^{\rm osc}\left(x\gg \gamma t(\eta^2-1)\right) = C^{\rm osc}_0(x)\left(\frac{\alpha}{2\gamma t (\eta^2-1)}\right)^{\delta-2}
\end{gather}
while in the steady state with $\alpha\ll x\ll \gamma t (\eta^2-1)$, it reads as
\begin{gather}
C^{\rm osc}(x\ll\gamma t(\eta^2-1)) = C^{\rm osc}_0(x)\left(\frac{\alpha}{x}\right)^{\delta-2}=\nonumber \\ =\frac{2\cos(2k_F x)}{(2\pi\alpha)^2}\left(\frac{\alpha}{x}\right)^{\delta}
\end{gather}
showing a non-trivial power-law decay. The exponent is calculated as
\begin{gather}
\delta = 2\frac{\eta^2+1}{\eta^2-1} - 4 \frac{\eta(\eta^2-1)}{\left(\frac{v}{\gamma}\right)^2+\left(\eta^2-1\right)^2}
\end{gather}
where $v$ is the Fermi velocity and $\gamma$ is the coupling to the environment. As discussed previously, $\eta$ cannot exceed $\eta_c$, therefore within the $1<\eta<\eta_c$ interval,
the relation $\delta>2$ holds.

\end{document}